\documentclass[aps,pre,twocolumn,nofootinbib,amsfonts,amssymb,amsmath,longbibliography]{revtex4-1}
\usepackage{amsthm}
\usepackage{graphicx,xcolor,soul}
\usepackage{bm}
\usepackage[colorlinks,linkcolor=magenta,citecolor=magenta]{hyperref}

\newcommand{\dd}{\mathrm{d}}
\newcommand{\pd}[2]{\frac{\partial #1}{\partial #2}}
\newcommand{\td}[2]{\frac{\dd #1}{\dd #2}}
\newcommand{\Int}[1]{\int\dd #1\;}
\newcommand{\IInt}[3]{\int_{#2}^{#3}\dd #1\;}
\newcommand{\mean}[1]{\langle #1\rangle}
\newcommand{\bset}[2]{\left\{#1\;|\;#2\right\}}

\newcommand{\al}{\alpha}
\newcommand{\gam}{\gamma}
\newcommand{\lam}{\lambda}
\newcommand{\sig}{\sigma}
\newcommand{\vhi}{\varphi}

\newcommand{\J}{\mathcal J}
\newcommand{\trj}{\bm\imath}
\newcommand{\f}{^\ast}

\begin{document}

\title{Modeling of biomolecular machines in non-equilibrium steady states}

\author{Thomas Speck}
\affiliation{Institut f\"ur Physik, Johannes Gutenberg-Universit\"at Mainz, Staudingerweg 7-9, 55128 Mainz, Germany}

\begin{abstract}
  Numerical computations have become a pillar of all modern quantitative sciences. Any computation involves modeling--even if often this step is not made explicit--and any model has to neglect details while still being physically accurate. Equilibrium statistical mechanics guides both the development of models and numerical methods for dynamics obeying detailed balance. For systems driven away from thermal equilibrium such a universal theoretical framework is missing. For a restricted class of driven systems governed by Markov dynamics and local detailed balance, stochastic thermodynamics has evolved to fill this gap and to provide fundamental constraints and guiding principles. The next step is to advance stochastic thermodynamics from simple model systems to complex systems with ten thousands or even millions degrees of freedom. Biomolecules operating in the presence of chemical gradients and mechanical forces are a prime example for this challenge. In this Perspective, we give an introduction to isothermal stochastic thermodynamics geared towards the systematic multiscale modeling of the conformational dynamics of biomolecular and synthetic machines, and we outline some of the open challenges.
\end{abstract}

\maketitle

%% ---- introduction ----

\section{Introduction}

Equilibrium statistical mechanics is concerned with configurations and their statistics, which requires as the only input a Hamiltonian assigning every configuration an energy~\cite{chandler}. These two components, the relevant degrees of freedom captured through the configuration and the Hamiltonian, have to be modeled on physical grounds. From a computational perspective, equilibrium statistical mechanics underlies virtually all existing approaches to model and simulate biomolecules and biomolecular systems~\cite{frenkel}. Moreover, statistical mechanics enables a wide range of advanced sampling approaches that have been developed over the last decades such as umbrella sampling~\cite{torrie77}, forward flux sampling~\cite{alle06,hussain20} and variants~\cite{erp05}, and transition path sampling~\cite{bolh02,dell02}. Typical tasks of simulations include revealing and sampling microscopic pathways (\emph{e.g.}, folding pathways of proteins and peptides~\cite{shaw11}), computing free energy differences~\cite{poho10,shirts12,mey20}, and computing reaction rates. In contrast, for driven systems the arsenal of numerical methods is still restricted and investigations often fall back to ``vanilla'' molecular dynamics simulations. But even these are confronted with, \emph{e.g.}, the choice of thermostats, which often lack a rigorous theoretical basis out of equilibrium.

Quite clearly do biomolecular machines like ribonucleases, motor proteins such as kinesin, and the various ATPases operate away from thermal equilibrium. Not only are biomolecular machines driven, they are also ``small'' and have to operate reliably in an aqueous crowded environment in which thermal fluctuations cannot be neglected. Invoking physical and thermodynamic arguments to understand the operation of these machines has a long history, starting with Schrödinger's ``What is life?''~\cite{schrodinger}. The work of T.L. Hill on the transduction of free energy has been highly influential~\cite{hill77}. Theoretical aspects of the modeling of molecular machines have been reviewed in Refs.~\citenum{juli97,bust01,kolo07}. Many of these aspects have also entered the development of stochastic thermodynamics~\cite{andr06,seif11}, which has evolved into a comprehensive theoretical framework over the past two decades~\cite{seif12,ciliberto17,horo19} (see Ref.~\cite{brown19} for an excellent and gentle introduction). Stochastic thermodynamics systematically extends thermodynamic notions like work, heat, entropy (even efficiency~\cite{verl14}) to individual stochastic trajectories (histories of configurations). It is now established that thinking about these trajectories within the mathematical framework of large deviations~\cite{touc09,touchette18,jack20} is an extremely powerful way to address small systems driven away from thermal equilibrium.

A recurring theme in modeling complex systems and materials is the need to coarse-grain, \emph{i.e.}, to reduce the degrees of freedom of the original (atomistic) model to make it amenable to numerical investigations~\cite{noid08,murtola09}. Not only are computational capacities limited (though ever-expanding), and therefore the accessible length and time scale, but often certain microscopic details are irrelevant for the question at hand. Coarse-graining methods can be roughly divided into two strategies, one motivated by polymers and one by proteins. The first strategy is \emph{structural} coarse-graining through combining a number of heavy atoms together with their associated hydrogens into units (called beads) and determining their effective interactions~\cite{izvekov05,noid08}, which often involves higher-body interactions~\cite{scherer18}. This route has been followed successfully for suspensions of small molecules and polymers, which can be decomposed into a few recurrent chemical motifs represented by these beads. The prototypical example is the MARTINI force field~\cite{marrink07}.

The alternative approach is to think in terms of molecular \emph{conformations}, dynamically distinguishable sets of structurally similar atomistic configurations. Conformations are stabilized by hydrogen bonds and collective (often hydrophobic) forces. They exhibit long dwell times and sudden transitions between conformations. Underlying this picture is a time-scale separation between the microscopic motion of atomic constituents and the collective reorganization into a different conformation. For dynamics obeying detailed balance, Markov state modeling~\cite{pand10,prin11,chodera14,husic18} has developed into a powerful computational framework that allows to systematically construct discrete models of conformational dynamics from atomistic molecular dynamics simulations and experimental data. Focusing on the conformational dynamics allows to access much longer timescales than possible in molecular dynamics simulations without having to make an \emph{a priori} choice on relevant structural features. Markov state modeling has been successfully applied to study the folding kinetics and pathways of small peptides~\cite{abella20}, which might be governed by atomistic details that become lost through structural coarse-graining~\cite{rudz16}.

Extending Markov state modeling to biomolecular machines requires to take properly into account mechanical forces and the conversion of chemicals, both breaking detailed balance and implying a steady discharge of heat into the environment. This is exactly the realm of stochastic thermodynamics, which so far, however, has mostly been used to study ``top-down'' models with a few, rather abstract, conformations. Here we attempt a synthetic review of ideas from stochastic thermodynamics with the perspective of applying it in the systematic ``bottom-up'' construction of discrete models describing the coarse-grained conformational dynamics of biomolecular machines. The route we take is through accounting for the exchanges of a machine with its environment. Starting from the familiar canonical ensemble in Sec.~\ref{sec:statmech}, we systematically develop a unified picture of dynamics and energetics of systems in contact with ideal reservoirs, first for currents close to equilibrium (Sec.~\ref{sec:lr}) and then more general (Sec.~\ref{sec:nlr}). We then review the mathematical tools necessary to treat fluctuations (Sec.~\ref{sec:fluct}) before sketching three challenges: transferability, coarse-graining, and numerical sampling methods (Sec.~\ref{sec:disc}).

%% ---- main ----

\section{Fluctuations and large deviations}
\label{sec:statmech}

We first consider conventional statistical mechanics of equilibrium systems coupled to reservoirs to establish some notions. What makes statistical mechanics so successful is the direct connection of \emph{statistical} partition functions with the corresponding \emph{thermodynamic} potentials, which depend only on a few thermodynamic variables such as temperature. These variables come in pairs, one extensive (\emph{e.g.} volume, energy, \dots) and one intensive (pressure, temperature, \dots), both of which are related through the thermodynamic potential. A change of variables (corresponding to another ensemble) is achieved through the Fenchel-Legendre transformation.

\begin{figure}[t]
  \centering
  \includegraphics{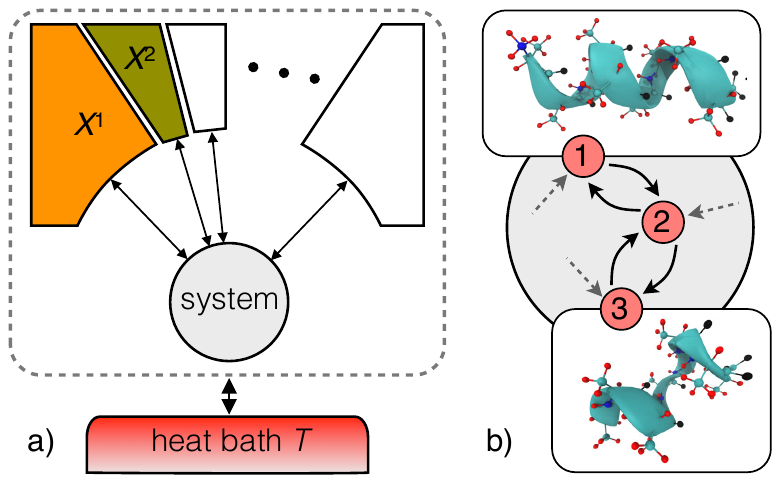}
  \caption{(a)~Sketch of the total system (dashed boundary) coupled to a heat bath. The total system is divided into a number of ideal reservoirs, each exchanging an extensive quantity $X^\al$ with the system proper (gray). (b)~The system is further divided into molecular conformations represented as a graph and endowed with a stochastic Markov dynamics. Snapshots show two conformations of deca-alanine, $\al$-helix (state 1) and an example for a misfolded intermediate (state 3)~\cite{knoch18}.}
  \label{fig:system}
\end{figure}

The extensive quantities can be divided further into conserved quantities $\bar X^\al$, meaning they are constant and thus parameters, and non-conserved quantities. The latter can be represented through dividing a total system into the system proper holding the amounts $\bar X^\al-X^\al$ and ``reservoirs'' with $X^\al$, which is sketched in Fig.~\ref{fig:system}(a). Diverting from the usual path, we focus on the reservoirs (the reason will become clear in the next section). For example, consider a volume $\bar V$ filled with a solvent and holding $\bar N$ identical, diffusing molecules. Now split the volume into a (small) system and a (large) reservoir. Clearly, the number $X^1=N$ of molecules within the reservoir's volume $V$ is a stochastic quantity since molecules enter and leave the system randomly. Hence, in general there is a probability $P(X;\bar X)$ with vector $X=(X^1,\dots)$ and $X_t$ denotes the specific random values at time $t$.

For our example, the probability $P(N)$ for non-interacting molecules is a binomial distribution. Exploiting Stirling's approximation, we find
\begin{equation}
  \label{eq:ld:ideal}
  \ln P(N;V) = -Vh(n) + \cdots
\end{equation}
where the additional terms are sublinear in $V$; and
\begin{multline}
  h(n) = -\frac{\bar n}{\gam}\ln\frac{\bar n}{\gam} + n[\ln n-\ln\gam] \\ + \left(\frac{\bar n}{\gam}-n\right)\left[\ln \left(\frac{\bar n}{\gam}-n\right)-\ln(1-\gam)\right]
\end{multline}
with $n\equiv N/V$ the number density, global density $\bar n\equiv\bar N/\bar V$, and $\gam\equiv V/\bar V$ is the fixed volume fraction occupied by the reservoir. Equation~\eqref{eq:ld:ideal} is called a \emph{large deviation principle} and we write $P(N;V)\asymp e^{-Vh(n)}$. The function $h(n)$ is called the \emph{rate function}. In statistical mechanics, the rate function is related to the thermodynamic potential, in this example to the (Gibbs) free energy $\mathcal G(N,V)=Vh(n)$. Throughout, we consider systems at constant temperature $T$ and measure energies in units of the thermal energy $k_\text{B}T$ and entropy in units of Boltzmann's constant $k_\text{B}$. The most likely value $\bar n$ of the density follows from the minimum, $\partial_nh=0$, with $h(n)$ describing the fluctuations of the density $n$ around $\bar n$.

Now let us assume that within the system we can fix the chemical potential of the molecules to $\mu$, effectively generating an inhomogeneous density that is different in system and reservoir. This modifies the probability to $P_\mu(N;V)=P(N;V)e^{\mu N}/Z$. Although here the partition function
\begin{equation}
  \label{eq:ld:Z}
  Z(\mu;V) = \sum_{N=0}^\infty P(N;V)e^{\mu N} \asymp e^{-V\omega(\mu)}
\end{equation}
was introduced as a simple normalization factor, it actually contains all the information about the system and arguably is the most important quantity in equilibrium statistical mechanics. In the second step, we have again assumed a large deviation principle, where rate function $h(n)$ and large deviation function $\omega(\mu)$ are related through the Fenchel-Legendre transform
\begin{equation}
  \label{eq:ld:flt}
  \omega(\mu) = \inf_n [h(n)-\mu n].
\end{equation}
Intuitively, this result can be understood rather easily: For $V\to\infty$ the integrand in Eq.~\eqref{eq:ld:Z} is dominated by a single value of $n$ (treated as a continuous variable), for which the argument $h-\mu n$ of the exponential function becomes the smallest. The infimum is attained for
\begin{equation}
  \mu = \pd{h}{n} = \pd{\mathcal G}{N},
\end{equation}
which simply states that the chemical potential in the reservoir is equal to the prescribed chemical potential of the system as expected for thermal equilibrium. We say that $\mu$ and $N$ are \emph{conjugate variables} with respect to the potential $\mathcal G$. Equation~\eqref{eq:ld:flt} is a variational principle minimizing the biased potential $h(n)-\mu n$.

While this excursion provides the minimal background we will need in the following, for further reading we recommend Ref.~\citenum{touc09} as an excellent introduction into the formalism of large deviations.

%% ---- linear response ----

\section{Linear response}
\label{sec:lr}

\subsection{Fluctuations of currents}
\label{sec:lr:curr}

Let us consider the following situation: We have no information about the internal state of our molecular machine (the system proper) but we can resolve the exchange of extensive quantities between the system and the reservoirs. This means we can determine the changes (also called generalized distances)
\begin{equation}
  \Delta_\tau \equiv X_\tau-X_0 = \IInt{t}{0}{\tau} \dot X_t
  \label{eq:Delta}
\end{equation}
over a prescribed time $\tau$. In the following, the dot denotes a rate (but not necessarily a total derivative with respect to time). In equilibrium, there is a free energy $\mathcal G(X)$ that depends on the state of the reservoirs. Moreover, random exchanges of $X$ occur also in equilibrium but on average they have to be zero,
\begin{equation}
  \mean{\Delta^\al_\tau}_\text{eq} = \Int{\Delta}\Delta^\al P_\text{eq}(\Delta;\tau) = 0,  
\end{equation}
since there can be no transport. This implies that the probability distribution of exchanges in equilibrium is symmetric, $P_\text{eq}(-\Delta)=P_\text{eq}(\Delta)$, and the most likely value is $\Delta=0$. According to the first law, any change of the free energy is due to heat $Q$ exchanged with the heat bath (at unit temperature). Per unit time we find
\begin{equation}
  \dot{Q} = -\td{\mathcal G}{t} = -\pd{\mathcal G}{X^\al}\dot X^\al = f_\al\dot X^\al
  \label{eq:diss}
\end{equation}
with affinities $f_\al\equiv-\pd{\mathcal G}{X^\al}$. Each $f_\al$ and $X^\al$ form a conjugate pair in analogy with $\mu$ and $N$. We follow Einstein's sum convention and sum over repeated greek indices. From now on we assume \emph{ideal reservoirs} for which the affinities $f=(f_1,\dots)$ are constant and input parameters to the theory. Integration over the time $\tau$ thus yields $Q_\tau=f_\al\Delta^\al_\tau$. Clearly, $\mean{Q_\tau}_\text{eq}=0$.

Even arbitrarily far from equilibrium, the fluctuations captured through the distribution $P(\Delta;\tau)$ are constrained by fluctuation theorems~\cite{gall95,jarz07,kurc98,lebo99,croo99,hata01,andr07a}. The physical picture is rather simple: Driving the system requires a non-vanishing dissipation and breaks time-reversal symmetry, which in steady state yields the fluctuation theorem
\begin{equation}
  \label{eq:ft}
  \frac{P(\Delta_\tau)}{P(-\Delta_\tau)} \asymp e^{Q_\tau} = e^{f_\al\Delta^\al_\tau}.
\end{equation}
It quantifies the likelihood of breaking time-reversal symmetry through the dissipated heat $Q_\tau$. A non-zero average $\mean{Q_\tau}>0$ means that transport occurs from at least one reservoir to another through the system.

\subsection{Onsager's principle}

While Eq.~\eqref{eq:ft} is a general result, we now focus on the regime close to equilibrium. We assume that the departure from the symmetric distribution $P_\text{eq}(\Delta)$ can be written (justified in appendix~\ref{sec:bias})
\begin{equation}
  P(\Delta;\tau) = \frac{P_\text{eq}(\Delta;\tau)e^{\frac{1}{2}f_\al\Delta^\al}}{Z(f;\tau)},
  \label{eq:onsa:P}
\end{equation}
which manifestly obeys Eq.~\eqref{eq:ft}. The function
\begin{equation}
  Z(f;\tau) = \Int{\Delta} P_\text{eq}(\Delta;\tau)e^{\frac{1}{2}f_\al\Delta^\al}
  \label{eq:onsa:Z}
\end{equation}
again ensures that $P(\Delta)$ is a normalized probability distribution. We further assume a large deviation principle for the decay of the probability
\begin{equation}
  P_\text{eq}(\Delta;\tau) \asymp e^{-\tau\Phi(\J)/2}, \quad Z(f;\tau) \asymp e^{-\tau\psi(f)/2}
\end{equation}
with (macroscopic) currents $\J^\al\equiv\Delta^\al/\tau$, but now the trajectory length $\tau$ takes the role of the large parameter. Plugging both expressions into Eq.~\eqref{eq:onsa:Z}, we find Onsager's principle~\cite{onsa31,onsa31a}
\begin{equation}
  \label{eq:onsa:psi}
  \psi(f) = \inf_\J[\Phi(\J)-f_\al\J^\al],
\end{equation}
which relates affinities and currents through a Fenchel-Legendre transform in analogy with Eq.~\eqref{eq:ld:flt}. We thus recover the same structure as for equilibrium statistical mechanics, whereby $\Phi(\J)$ and $\psi(f)$ play the role of thermodynamic potentials for currents. This analogy can be exploited to obtain transport coefficients~\cite{palm17}.

\subsection{Some consequences}

Let us expand Eq.~\eqref{eq:onsa:P} to linear order of $Q$, which yields the average $\mean{\Delta^\al_\tau}=\tau\chi^{\al\beta} f_\beta$ with conductivities
\begin{equation}
  \chi^{\al\beta} = -\left.\pd{^2\psi}{f_\al\partial f_\beta}\right|_\text{eq} = \lim_{\tau\to\infty}\frac{1}{2\tau}\mean{\Delta^\al_\tau\Delta^\beta_\tau}_\text{eq} = \pd{\mean{\mathcal J^\al}}{f_\beta}.
  \label{eq:chi}
\end{equation}
Since we can change the order of partial derivatives, the matrix $\chi$ is symmetric, which is often referred to as Onsager reciprocal relations. The equilibrium fluctuations of the quantities exchanged with the reservoirs thus determine their non-equilibrium averages. Eq.~\eqref{eq:chi} is called the fluctuation-dissipation theorem. The average dissipation reads
\begin{equation}
  \mean{Q_\tau} = \tau f_\al\chi^{\al\beta}f_\beta > 0,
\end{equation}
from which we conclude that $\chi^{\al\beta}$ is a positive semi-definite matrix\footnote{Note the nice link to geometry~\cite{weinhold75,ruppeiner79}. As implied by our notation, extensive quantities can be seen as (contravariant) components of a vector $X$ while the affinities $f$ form a covector. The matrix $\chi^{\al\beta}$ induces a metric in this space which endows dissipation with the interpretation of a length~\cite{feng08,siva12}.}. The inverse matrix $\chi_{\al\beta}$ (resistivities) determining affinities through $f_\al=\chi_{\al\beta}\J^\beta$ follows from $\chi^{\al\gam}\chi_{\gam\beta}=\delta^\al_\beta$.

Note that we can recast Eq.~\eqref{eq:onsa:psi} into the variational problem $\inf_{\dot X}\mathcal R(\dot X)$ with respect to the rates $\dot X$ with function (see appendix~\ref{sec:ray})
\begin{equation}
  \mathcal R(\dot X) \equiv \frac{1}{2}\dot X^\al\chi_{\al\beta}\dot X^\beta + \td{\mathcal G}{t},
  \label{eq:ray}
\end{equation}
which is sometimes called the ``Rayleighian''~\cite{zhou18}. This form more clearly exposes the formal similarities with the Lagrangian in mechanics with a quadratic ``mass'' term and $-\td{\mathcal G}{t}$ taking the role of the potential energy.

%% ---- beyond ----

\section{Beyond linear response}
\label{sec:nlr}

\subsection{Discrete state space}

To go further away from equilibrium, we need to be more specific about the internal working of the system and in particular its coupling to the reservoirs. Full information is contained in the microstate $\xi$ comprising positions and momenta of all atoms. Practically, it is neither possible nor necessary to fully resolve microstates, which, \emph{e.g.}, for a solvated colloidal particle would contain information about all solvent molecules. We assume that meaningful mesostates $\{i\}$ can be defined, sets of microstates representing for example conformations of a molecule [cf. Fig.~\ref{fig:system}(b)] or the (discrete) position of a colloidal particle. Moreover, we will assume that the dynamics of these mesostates decouples from the much faster dynamics of microstates yielding a Markov process with stochastic transitions between the mesostates. Such a discrete representation is either obtained ``top-down'' through general considerations, but can also be constructed systematically, for example through Markov state modeling based on molecular dynamics simulations~\cite{prin11}. The importance of this step and the challenges of dimensionality reduction cannot be overemphasized~\cite{sittel18}. However, it is not the focus of this perspective and we will assume in the following that a discrete state space has been constructed.

To make the connection with the discussion so far, we have to realize that during every transition $i\to j$ between mesostates there might be corresponding changes in the environment so that the quantity $X^\al$ held by the reservoir changes as
\begin{equation}
  X^\al \to X^\al + d^\al_{ij}.
\end{equation}
Clearly, the couplings $d^\al_{ij}$ are antisymmetric, $d^\al_{ji}=-d^\al_{ij}$, since going back undoes the change. Table~\ref{tab:coup} summarizes the extensive quantities $X$ and affinities $f$ of interest in the present context.

%% ---- table of couplings ----
\begin{table}[b!]
  \begin{tabular}{l|l|l}
    \hline\hline
    extensive $X$ & intensive affinity $f$ & $\chi$ \\
    \hline
    electric charge $Q_\text{el}$ & electric potential $-\phi_\text{el}$ & conductivity \\
    number of molecules $N$ & chemical potential $-\mu$ & mobility \\
    traveled distance $x$ & force $-f_L$ & mobility \\
    angle $\vhi$ & torque & rot. mobility \\
    strain & stress & viscosity \\
    \hline\hline
  \end{tabular}
  \caption{Summary of the most relevant reservoir types and their corresponding affinities.}
  \label{tab:coup}
\end{table}

\subsection{Illustration: Enzyme}
\label{sec:enzyme}

\begin{figure}[t]
  \centering
  \includegraphics{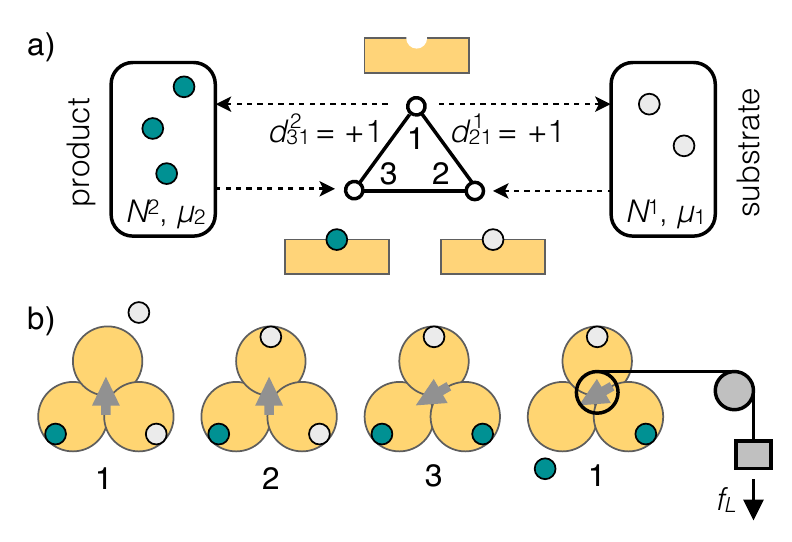}
  \caption{(a)~Sketch of an enzyme with three internal states: unbound (1), bound substrate (2), and bound product (3). Two transitions ($1\leftrightarrow2$ and $1\leftrightarrow3$) exchange molecules with the reservoirs while $2\leftrightarrow3$ converts bound molecules between substrate and product. (b)~The same internal states applied to F$_1$-ATPase, a rotary molecular motor consisting of three identical subunits and a central shaft (arrow). The substrate molecule is now ATP, which is hydrolyzed to ADP and phosphate (product), releasing the free energy $\mu_1-\mu_2$. Part of this free energy is used to rotate the central shaft in the transition $2\leftrightarrow3$, returning the motor to the unbound state (1) but with the shaft rotated by 120$^\circ$. The last sketch shows the coupling to an ideal work reservoir represented as a weight that is lifted against the force $f_L$.}
  \label{fig:enzyme}
\end{figure}

As the canonical example, Fig.~\ref{fig:enzyme}(a) shows a sketch of an enzyme (the system proper) converting chemical species 1 (the substrate) into species 2 (the product)~\cite{seifert18}. On the coarsest level, the enzyme is modeled with three internal states: unbound (1), bound substrate (2), and bound product (3). It is coupled to two reservoirs holding $N^\al$ molecules, and we assume that the corresponding chemical potentials $\mu_\al=-f_\al$ are functions of temperature and pressure alone and held constant. Note that the two reservoirs are not spatially separated but \emph{kinetically} separated, with the conversion between product and substrate so slow outside the enzyme that it is neglected. In the transition $1\to2$ a substrate molecule is bound to the enzyme and thus $d^1_{12}=-1$, whereby the substrate molecule is return to the reservoir in $2\to1$ with $d^1_{21}=1$\footnote{A subtle but important point is that in our framework there are no separate rates for product (substrate) to attach to (detach from) the enzyme, but transitions of the enzyme simultaneously involve the uptake or release of a molecule.}. The transition $2\to3$ converts the bound substrate into bound product, which is released in the transition $3\to1$.

Since the total number of molecules is conserved, $\bar N=N^1+N^2$, we find the Gibbs free energy
\begin{equation}
  \mathcal G(N^1,N^2) = -f_\al N^\al = \mu_1\bar N - (\mu_1-\mu_2)N^2.
\end{equation}
Requiring that the average dissipation rate $\mean{\dot Q}=(\mu_1-\mu_2)\mean{\dot N^2}$ is positive, we see that for $\mu_1>\mu_2$ substrate molecules are being converted to product on average. There is thus a macroscopic current from one reservoir to the other which is enabled by the enzyme although the driving force is the difference of chemical potentials. We assume that, even though reservoirs are depleted and filled, they are so big that the chemical potential can be treated as constant over sufficiently long times\footnote{To assess this assumption, let the product/substrate molecules form an ideal gas with chemical potential $\mu=\ln\lam^3N/V$, where $\lam$ is the thermal de Broglie wavelength. For any finite fixed reservoir volume $V$ the chemical potential will change by $\ln(1\pm1/N)\approx\pm1/N$ upon removal/addition of a molecule.}, during which a steady state ensues.

Things become interesting if a transition involves a conformational change that exerts work on the environment, a \emph{power stroke}. Consider for example a rotary molecular motor such as F$_1$-ATPase, a paradigmatic molecular machine that has been studied extensively~\cite{noji97,wang98a}. This motor rotates in 120$^\circ$ steps separated by dwells, hydrolyzing one ATP molecule in each stroke. A coarse sketch of the involved conformations is shown in Fig.~\ref{fig:enzyme}(b), where now the transition $2\leftrightarrow3$ involves a conformational change that rotates the shaft. This rotation can be accounted for as a change in the environment with $d^3_{23}=2\pi/3$, where $X^3=\vhi$ is the total angle the shaft has rotated a spool with radius $R$. We can attach a load represented as a weight that is lifted against the force $f_L$, the potential energy of which is $f_LR\vhi$. The total free energy now reads
\begin{equation}
  \mathcal G(N^1,N^2,\vhi) = \mu_1\bar N - (\mu_1-\mu_2)N^2 + f_LR\vhi,
\end{equation}
from which we read off the affinity $f_3=-f_LR$, \emph{i.e.}, the torque applied to the shaft. Note that microscopic reversibility implies that when performing work \emph{on} the molecular motor (lowering the weight) it is rotating in the opposite direction synthesizing ATP molecules, which is indeed observed~\cite{itoh04}.

\subsection{Local detailed balance}
\label{sec:ldb}

To continue we need to know the evolution equation for the mesostates. Before, however, we ask ourselves how the dynamics of the enzyme is affected by the driving. To answer this question, we first go back to thermal equilibrium, in which the probability to find the system in mesostate $i$ is\footnote{We are economical and use the same symbol $i$ for the index of the mesostate and the set of microstates comprising the mesostate.}
\begin{equation}
  p_i(X) = \sum_{\xi\in i} P(\xi;X) = e^{-[G_i(X)-\mathcal G(X)]}
\end{equation}
with microscopic joint probability
\begin{equation}
  \label{eq:boltz}
  P(\xi;X) = e^{-[\mathcal H(\xi)-\mathcal G(X)]}
\end{equation}
given by the Boltzmann weight, where $\mathcal H(\xi)$ is the Hamiltonian. Here we have to distinguish the global free energy $\mathcal G(X)$ from the ``constrained'' free energies $G_i(X)$ of the mesostates.

\begin{figure}[t]
  \centering
  \includegraphics{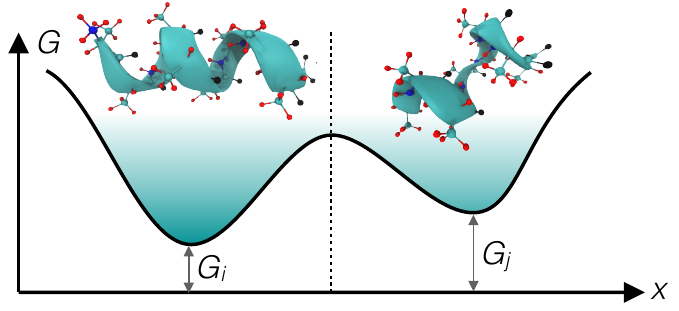}
  \caption{Sketch of a free energy landscape $G(x)$ along some continuous reaction coordinate $x$ capturing the slowest mode. Mesostates correspond to ``wells'', the vicinity of minima with free energies $G_i$ separated by barriers. Relaxation of all other degrees of freedom is faster than barrier crossing so that the distribution of microstates approaches the Boltzmann distribution Eq.~\eqref{eq:boltz} on time scales (much) shorter than the lifetime of mesostates.}
  \label{fig:timesep}
\end{figure}

Most importantly, we assume a timescale separation between the relaxation of microstates within mesostates, and transitions between mesostates, cf. Fig.~\ref{fig:timesep}. Since mesostates are a coarse-grained representation, their dynamics is necessarily stochastic and determined by the time-independent non-negative transition rates $w_{ij}$ for each transition $i\to j$. We will call the set of rates $w$ a \emph{Markov model}. Since the composite system is only coupled to a heat reservoir but otherwise closed, these transition rates obey the \emph{detailed balance} condition
\begin{equation}
  \label{eq:db}
  \frac{w_{ij}}{w_{ji}} = \frac{p_j}{p_i} = e^{-(G_j-G_i)},
\end{equation}
which guarantees the absence of dissipation.

Making the changes in the reservoirs explicit, we have
\begin{multline}
  \label{eq:ldb:pre}
  \ln\frac{w_{ij}(\{X^\al\to X^\al+d^\al_{ij}\})}{w_{ji}(\{X^\al+d^\al_{ij}\to X^\al\})} \\ = -\left[G_j(\{X^\al+d^\al_{ij}\})-G_i(\{X^\al\})\right]
\end{multline}
after taking the logarithm. We again appeal to the concept of ideal reservoirs, from which follow additive free energies $G_i(X)=G^0_i-f_\al X^\al$ with bare free energies $G^0_i$ independent of $X$. Plugging this sum back into Eq.~\eqref{eq:ldb:pre}, we obtain 
\begin{equation}
  \label{eq:ldb}
  \ln\frac{w_{ij}}{w_{ji}} = -(G^0_j-G^0_i) + f_\al d^\al_{ij},
\end{equation}
which is known as the \emph{local detailed balance} condition\footnote{The concept of local detailed balance seems to have appeared first in Ref.~\citenum{bergmann55} by Bergmann and Lebowitz. This paper has been revisited recently with a different focus from this Perspective~\cite{klein21}.}. The right hand side manifestly is independent of $X$, which implies that also the transition rates $w_{ij}$ are independent of $X$ and we do not have to keep track of the absolute values of the environmental variables.

In this setting, non-equilibrium is achieved by preparing an initial state that exhibits currents between reservoirs (trying to reach equilibrium) which are assumed to persist for a sufficiently long time. While the joint distribution $P(\xi,X)$ diverts from the Boltzmann factor [Eq.~\eqref{eq:boltz}], the dynamics of the combined system-reservoirs still obeys detailed balance. These currents necessarily influence the stochastic dynamics of the system proper as captured by the local detailed balance condition Eq.~\eqref{eq:ldb}.

\subsection{Markov dynamics and graphs}

The stochastic dynamics of the system (in continuous time) is determined by the time-independent non-negative transition rates $w_{ij}$. The probability $p_i(t)$ to find the system in mesostate $i$ evolves according to the master equation
\begin{equation}
  \label{eq:master:time}
  \partial_tp_i(t) = \sum_{j\neq i} \left[p_j(t)w_{ji} - p_i(t)w_{ij}\right].
\end{equation}
It is helpful to interpret the discrete state space as a graph in which the mesostates correspond to the vertices, $V\equiv\{i\}$, and possible transitions $E\equiv\{i\leftrightarrow j\}$ form the edges of the graph [Fig.~\ref{fig:system}(b)]. We require two properties: (i)~the graph is connected and (ii)~for every $w_{ij}>0$ also $w_{ji}>0$ (\emph{i.e.}, every edge can be traversed in both directions). The first property means that any two vertices of the graph are connected by a path. The second property is called \emph{microscopic reversibility} and guarantees the absence of absorbing states. These properties are sufficient to proof existence and uniqueness of a stationary solution, \emph{i.e.}, the solution $p$ of
\begin{equation}
  \label{eq:master}
  \sum_{j\neq i} [p_jw_{ji} - p_iw_{ij}] = 0
\end{equation}
is reached from any initial distribution after a transient. The proof can be found, \emph{e.g.}, in Ref.~\citenum{schn76}.

In the following, we will focus on this steady state with probabilities $p$. We define the non-negative \emph{fluxes}
\begin{equation}
  \label{eq:phi}
  \phi_{ij} \equiv p_iw_{ij}
\end{equation}
and with these the probability \emph{currents} and \emph{affinities}
\begin{equation}
  \label{eq:aff}
  j_{ij} \equiv \phi_{ij} - \phi_{ji}, \qquad
  a_{ij} \equiv \ln\frac{\phi_{ij}}{\phi_{ji}},
\end{equation}
respectively. The case where all currents vanish individually with $\phi_{ji}=\phi_{ij}$ corresponds to detailed balance [it immediately leads to Eq.~\eqref{eq:db}], and the system is in thermal equilibrium. If detailed balance is broken then probability currents and affinities along (at least some) edges are non-zero, and probabilities $p_i\neq e^{-G^0_i}$ are no longer given by Boltzmann weights. Table~\ref{tab:sym} summarizes most of the symbols introduced so far.

\begin{table}[t]
  \begin{tabular}{l|c|p{.6\linewidth}}
    \hline\hline
    macroscopic & $X^\al$ & quantities in the environment (reservoirs) that are influenced by the system \\
    & $\Delta^\al$, $\J^\al$ & their changes and currents \\
    \hline
    mesoscopic & $d^\al_{ij}$ & change of quantity $X^\al$ due to transition $i\to j$ of the system \\
    & $w_{ij}$ & transition rates \\
    & $p_i$ & probability of mesostate $i$ \\
    & $\phi_{ij}$, $j_{ij}$ & probability fluxes and currents within system \\
    \hline
    atomistic & $\xi$ & microstate \\
    & $\mathcal H$ & the Hamiltonian \\
    \hline\hline
  \end{tabular}
  \caption{Summary of the most important symbols.}
  \label{tab:sym}
\end{table}

We are concerned with two types of observables, those with one index ($A_i$) assigning every mesostate (vertex) a value, and those with two indices ($B_{ij}$) assigning edges a value. Expectation values for the former are obtained as $\mean{A_i}=\sum_iA_ip_i$ and for the latter $\mean{B_{ij}}=\sum_{ij}B_{ij}\phi_{ij}$, where the sum is over all mesostates $i$ and $j$ and thus includes every edge twice ($i\to j$ and $j\to i$). Clearly, $\mean{B_{ij}}$ is a rate. For antisymmetric edge observables with $B_{ji}=-B_{ij}$ one finds
\begin{equation}
  \mean{B_{ij}} = \sum_{ij}B_{ij}\phi_{ij} = \frac{1}{2}\sum_{ij}B_{ij}j_{ij}
\end{equation}
with probability currents $j_{ij}=\phi_{ij}-\phi_{ji}$. A very useful relation is Jensen's inequality
\begin{equation}
  \label{eq:jensen}
  \mean{g(A_i)} \geqslant g(\mean{A_i})
\end{equation}
holding for any convex real function $g(x)$.

\subsection{Cycles}

Currents and affinities are antisymmetric by construction. Due to this property it is useful to turn the graph into a directed graph $\vec G=(V,E,\nabla)$ by giving every edge $e\in E$ an orientation, which is described by the \emph{incidence matrix} $\nabla$ with entries
\begin{equation}
  \label{eq:nabla}
  \nabla_i^e \equiv
  \begin{cases}
    -1 & \text{if edge $e$ enters vertex $i$} \\
    +1 & \text{if edge $e$ leaves vertex $i$} \\
    0 & \text{otherwise (no relation)}
  \end{cases}
\end{equation}
having $|V|$ rows and $|E|$ columns~\cite{pole15}. The master equation~\eqref{eq:master} can then be written
\begin{equation}
  \label{eq:kirch}
  \sum_{j\neq i} j_{ij} = \sum_{e\in E} \nabla_i^e j_e = 0
\end{equation}
or $\nabla\cdot\vec\jmath=0$, where $j_e\equiv j_{ij}$ is the current along the oriented edge $e=i\to j$ and $\vec\jmath=(j_1,\dots,j_{|E|})$ is the vector of all edge currents. Eq.~(\ref{eq:kirch}) is also known as Kirchhoff's current law and expresses the conservation of probability through demanding that the total current into a vertex equals the current out of that vertex. Note that the actual orientation of edges does not matter, it merely gives ``forward'' and ``backward'' a meaning but we can traverse all edges in both directions. The null space (kernel) of the incidence matrix $\nabla$ is the space of directed simple \emph{cycles} in the graph $\vec G$~\cite{pole15}, see appendix~\ref{sec:cyc}.

An intuitive consequence of Kirchhoff's law is that in a steady state with time-independent probabilities $p$, all probability currents have to flow in cycles. Cycles set non-equilibrium apart from thermal equilibrium and enable transport, \emph{i.e.}, non-vanishing macroscopic currents $\mean{\J^\al}=\mean{d^\al_{ij}}\neq 0$. This might become clearer when again looking at the enzyme in Fig.~\ref{fig:enzyme}(a), which forms a single cycle. Traversing the cycle converts substrate into product (or vice versa) while returning the enzyme to its initial state.

\subsection{Stochastic energetics}

We finish this section by looking at the energetics of the system, which is at the very heart of stochastic thermodynamics. To account for the fact that mesostates represent finite volumes in phase space~\cite{seif19}, the bare free energy $G^0_i=U_i-S_i$ of mesostates is further split into internal energy
\begin{equation}
  U_i \equiv \sum_{\xi\in i} \mathcal H(\xi)P(\xi|i)
\end{equation}
and ``intrinsic'' entropy
\begin{equation}
  S_i \equiv -\sum_{\xi\in i}P(\xi|i)\ln P(\xi|i),
\end{equation}
respectively, with conditional probability $P(\xi|i)=P(\xi)/p_i$ for the system to be in microstate $\xi$ given that it resides in mesostate $i$. While the probabilities of mesostates divert from the Boltzmann factor, the local weight $P(\xi|i)=e^{-[\mathcal H(\xi)-G^0_i]}$ of microstates is assumed to follow the Boltzmann weight in line with the posited scale separation.

Having separated energy and intrinisc entropy of mesostates, the total change of entropy in a single transition $i\to j$ is composed of three contributions,
\begin{equation}
  \label{eq:S:tot}
  \underbrace{\delta Q_{ij} + (S_j-S_i)}_{\ln(w_{ij}/w_{ji})} - \ln\frac{p_j}{p_i} = \ln\frac{\phi_{ij}}{\phi_{ji}} = a_{ij},
\end{equation}
where $\delta Q_{ij}$ is the heat exchanged with the heat reservoir, the second term is the change of intrinsic entropy, and the last term is the change of (stochastic) entropy associated with the mesostates~\cite{seif05a,seif19}. To recover the established expression for the average entropy production rate
\begin{equation}
  \label{eq:sig}
  \dot\sig = \sum_{ij} \phi_{ij}\ln\frac{\phi_{ij}}{\phi_{ji}} = \mean{a_{ij}} \geqslant 0
\end{equation}
of Markov processes~\cite{schn76,lebo99,seif05a}, we identify the first two terms in Eq.~\eqref{eq:S:tot} with the transition rates. Combining this identification of the ratio $w_{ij}/w_{ji}$ with the local detailed balance condition Eq.~\eqref{eq:ldb}, we obtain the balance equation
\begin{equation}
  \label{eq:Q}
  \delta Q_{ij} + (U_j-U_i) = f_\al d^\al_{ij}
\end{equation}
for the energy exchanged in a transition between mesostates. This result avails itself of the interpretation as the first law $\delta U=\delta W-\delta Q$ across a single transition (the sign of heat is convention), from which we read off the work $\delta W_{ij}=f_\al d^\al_{ij}$ performed by the reservoirs on the system. For the three-state enzyme of Sec.~\ref{sec:enzyme} we find the work
\begin{equation}
  W_\circlearrowright = \delta W_{12} + \delta W_{23} + \delta W_{31} = \mu_1 - \mu_2 - f_LR\frac{2\pi}{3}
\end{equation}
for completing a single forward cycle.

%% ---- fluctuations ----

\section{Fluctuations}
\label{sec:fluct}

\subsection{Trajectory observables}

So far we have fixed the stochastic dynamics of mesostates, which is sufficient to calculate average rates and the statistics of state observables ($A_i$). However, it is not yet clear how to treat the fluctuations of currents and related time-extensive quantities. We now change our perspective and focus on the stochastic \emph{trajectories}, \emph{i.e.}, time-ordered sequences
\begin{equation}
  \trj \equiv (i_0,0)\to(i_1,t_1)\to\cdots\to(i_K,t_K)
\end{equation}
of mesostates visited within a fixed time $\tau$, whereby mesostates $i_\nu$, transition times $0<t_\nu<\tau$, and the number of transitions $K$ are random numbers.

Along trajectories we measure trajectory observables like currents, heat, work, etc.; all of which can be written as a functional (exemplified here for the heat)
\begin{equation}
  Q_\tau[\trj] = \sum_{\nu=1}^K \delta Q_{i_{\nu-1}i_\nu}
\end{equation}
summing the contributions of the antisymmetric edge observable $\delta Q_{ij}$ for every transition in the sequence $\trj$. Clearly, $Q_\tau$ is a random quantity with a probability distribution $P(Q;\tau)$.

We can now come back to our starting point looking at extensive quantities $X$ that are exchanged between the system and reservoirs. Tracing the trajectory of the system's mesostates together with the knowledge of the couplings $d^\al_{ij}$ allows to reconstruct the changes
\begin{equation}
  \Delta^\al_\tau[\trj] = \sum_{\nu=1}^K d^\al_{i_{\nu-1}i_\nu}
\end{equation}
within the reservoirs. Together with Eq.~\eqref{eq:Q} we thus obtain the heat $Q_\tau[\trj]=-(U_{i_K}-U_{i_0})+f_\al\Delta^\al_\tau[\trj]$ dissipated along a single trajectory. The term for the internal energy is a telescoping sum that reduces to a temporal boundary term, which vanishes when averaging over the subset of trajectories compatible with the given changes $\Delta_\tau$.

\subsection{Time-reversal symmetry}

Once we have obtained the rates $w$ of the Markov model it is straightforward to determine the probability of a single trajectory,
\begin{equation}
  \label{eq:traj}
  \mathcal P_\tau[\trj] = p_{i_0}e^{-r_{i_0}t_1} \prod_{\nu=1}^K w_{i_{\nu-1}i_\nu} e^{-r_{i_\nu}(t_{\nu+1}-t_\nu)},
\end{equation}
with $t_{K+1}=\tau$ and exit rates $r_i\equiv\sum_{j\neq i}w_{ij}$. Reading from left to right it says: probability to be in the initial mesostate $i_0$ times the probability to survive for time $t_1$ times the transition probability to a new mesostate $i_1$ times the survival probability for time $t_2-t_1$, and so on. Summing over all trajectories, $\sum_{\trj}\mathcal P_\tau[\trj]=1$.

Now suppose we ask for the probability $\mathcal P_\tau[\tilde\trj]$ to observe the reversed trajectory $\tilde\trj$, \emph{i.e.}, the same sequence of mesostates but traversed in the opposite direction. In thermal equilibrium, we could not discern whether a movie is played forward or backward and thus $\mathcal P_\tau[\trj]=\mathcal P_\tau[\tilde\trj]$. The ratio of these probabilities thus tells us how strongly this \emph{time-reversal symmetry} is broken due to the driving. It turns out that the logarithm of the ratio
\begin{equation}
  \label{eq:trs}
  \ln\frac{\mathcal P_\tau[\trj]}{\mathcal P_\tau[\tilde\trj]} = \sum_{\nu=1}^K\ln\frac{w_{i_{\nu-1}i_\nu}}{w_{i_\nu i_{\nu-1}}} + \ln\frac{p_{i_0}}{p_{i_K}} = \sum_{\nu=1}^K a_{i_{\nu-1}i_\nu} = \mathcal S_\tau[\trj]
\end{equation}
equals the sum of the affinities along the trajectory since the survival probabilities cancel (the sojourn times are the same for forward and backward trajectory), and we are left with the product of the transition rates in the first term. Plugging in Eq.~\eqref{eq:S:tot} followed by the first law Eq.~\eqref{eq:Q} yields the expression
\begin{equation}
  \mathcal S_\tau[\trj] = f_\al\Delta^\al_\tau[\trj] - (\Psi_{i_K}-\Psi_{i_0})
\end{equation}
for the total entropy produced along single trajectories with a temporal boundary term given by the effective potential $\Psi_i\equiv G^0_i+\ln p_i$. The agreement of the path entropy $\mathcal S_\tau\asymp Q_\tau$ with the actual dissipated heat due to exchanges with the reservoirs is what makes stochastic thermodynamics so powerful since it connects the second with the first law of thermodynamics on the level of stochastic trajectories.

Here is an example what we can do with Eq.~\eqref{eq:trs}. Let us look at the distribution of exchanges, $P(\Delta;\tau)$, which can be written
\begin{align*}
  P(\Delta) &= \sum_{\trj} \delta(\{\Delta^\al-\Delta^\al_\tau[\trj]\})\mathcal P[\trj] \\ &= \sum_{\tilde\trj}\delta(\{\Delta^\al+\Delta^\al_\tau[\tilde\trj]\})e^{\mathcal S}\mathcal P[\tilde\trj] \asymp P(-\Delta)e^{f_\al\Delta^\al}.
\end{align*}
In the first step, we have replaced $\Delta_\tau[\trj]=-\Delta_\tau[\tilde\trj]$ and used Eq.~\eqref{eq:trs}. We can then pull the exponential factor $e^{f_\al\Delta^\al}$ in front of the sum, which reduces to the distribution $P(-\Delta;\tau)$ for large $\tau$ (such that bounded boundary terms become negligble). We have thus derived the relation Eq.~\eqref{eq:ft} used previously, which restricts possible fluctuations of the currents. Note that instead of inverting the order of mesostates more general transformations can be considered, leading to a variety of fluctuation theorems. For a comprehensive discussion, see Seifert's review~\cite{seif12}.

Intriguingly, the entropy production also bounds fluctuations out of equilibrium. Using only the fluctuation theorem, Hasegawa and Vu~\cite{hasegawa19} obtained the following lower bound for the Fano factor
\begin{equation}
  F^\al_\tau \equiv \frac{\mean{(\Delta^\al_\tau)^2}-\mean{\Delta^\al_\tau}^2}{\mean{\Delta^\al_\tau}^2} \geqslant \frac{2}{e^{\tau\dot\sig}-1},
\end{equation}
measuring the dispersion (the variance divided by the squared mean) of the changes $\Delta_\tau$. Previously, a tighter bound $F^\al_\tau\geqslant 2/(\tau\dot\sig)$ holding for continuous-time Markov processes (as considered in this Perspective) has been proposed~\cite{bara15} and subsequently proven~\cite{ging16}. Such bounds have been termed thermodynamic uncertainty relations~\cite{pietzonka17,dechant18,koyuk20,falasco20,hartich21} and demonstrate that precision has a thermodynamic cost, \emph{i.e.}, reducing the dispersion of any process generally requires a larger dissipation with ramifications for biomolecular machines and processes like DNA replication.

\subsection{Empirical densities and fluxes}

Let us come back to the basic task of Markov State modeling: given an atomistic model and a discrete state space $\{i\}$ of conformations, what are the rates $w_{ij}$ so that the Markov evolution of mesostates follows the time evolution of the atomistic system? To tackle this question, let us assume that we can perform a number of (long) molecular dynamics simulations of the atomistic model, sampling trajectories $\xi_t$ of length $\tau$. Along a single finite trajectory $\xi_t$ of microstates with mapped sequence $\trj$ of mesostates, we measure the \emph{empirical} populations
\begin{equation}
  \hat p_i[\trj] \equiv \frac{1}{\tau}\IInt{t}{0}{\tau} \chi_i(\xi_t)
\end{equation}
and empirical fluxes ($\delta t\to 0$)
\begin{equation}
  \hat\phi_{ij}[\trj]\delta t \equiv \frac{1}{\tau}\IInt{t}{0}{\tau} 
  \chi_i(\xi_t)\chi_j(\xi_{t+\delta t})
\end{equation}
through counting. Here, $\chi_i(\xi)$ is an indicator function that is one if the microstate $\xi$ belongs to mesostate $i$ and zero otherwise. In the limit $\tau\to\infty$ we have that $\hat p_i\to p_i$, but for finite $\tau$, $\hat p$ and $\hat\phi$ are random quantities with a probability density.

The probability of these empirical quantities can be found through comparing the trajectory probability density of the (unknown) generating Markov model $w$ with those of a fictitious Markov model~\cite{maes08}. Skipping the technical details, the joint probability density
\begin{equation}
  P_\tau(\hat p=p\f,\hat\phi=\phi\f) \asymp e^{-\tau \mathcal I_w(p\f,\phi\f)}
\end{equation}
obeys a large deviation principle with rate function
\begin{equation}
  \label{eq:rf}
  \mathcal I_w(p\f,\phi\f) = \sum_{ij} \left[ 
    \phi\f_{ij}\ln\frac{\phi\f_{ij}}{p\f_iw_{ij}} - \phi\f_{ij} + p\f_iw_{ij} \right].
\end{equation}
Notable, the rate function is independent of the actual steady state probabilities $p$ and only the transition rates $w$ enter the rate function. Put differently, $P_\tau$ is the probability density to observe trajectories with empirical populations and fluxes that, in the limit $\tau\to\infty$, are typical for a fictitious model with transition rates $w\f_{ij}=\phi\f_{ij}/p\f_i$. As shown in appendix~\ref{sec:kl}, the rate function $\mathcal I_w$ equals the Kullback-Leibler divergence between the underlying process $w$ and the observed empirical densities and fluxes of a single trajectory.

\subsection{Estimating rates: The contraction principle}

The contraction principle allows to derive the rate function $I(y)$ for another variable $y=g(x)$ from a known rate function $I(x)$,
\begin{equation}
  I(y) = \inf_x\bset{I(x)}{y=g(x)}.
\end{equation}
This result expresses a simple fact: among rare events that with the largest probability (even though very small) will occur almost certainly. For example, the rate function for the probability currents reads
\begin{equation}
  I(p\f,j\f) = \inf_{\phi\f}\bset{\mathcal I_w(p\f,\phi\f)}{\phi\f_{ij}-\phi\f_{ji}=j\f_{ij}},
\end{equation}
\emph{i.e.}, the probability of currents is determined by the most likely realization of the fluxes obeying the constraints.

Here is another example: Assume we have harvested, in equilibrium, a single atomistic trajectory of length $\tau$ with populations $\hat p$ and fluxes $\hat\phi$. Given this limited data, what is the optimal Markov model $w$ with probabilities $p=\hat p$ and fluxes $\phi_{ij}=\hat p_iw_{ij}$? Exploiting Bayes' theorem, the probability that the Markov model $w$ has generated the data is $P(\phi|\hat\phi)\propto P_\tau(p,\hat\phi)$. To maximize this probability (called likelihood in this context), we invoke the contraction principle enforcing detailed balance as constraints on every edge,
\begin{equation}
  \inf_\phi\bset{\mathcal I_w(p,\hat\phi)}{\phi_{ij}-\phi_{ji}=0}.
\end{equation}
The constraints can be incorporated through Lagrange multipliers (the antisymmetry of the constraints implies antisymmetric Lagrange multipliers, $\lam_{ji}=-\lam_{ij}$),
\begin{equation}
  \pd{}{\phi_{ij}}\left[\mathcal I_w-\frac{1}{2}\sum_{kl}\lam_{kl}(\phi_{kl}-\phi_{lk})\right] = 0,
\end{equation}
which is solved by $\phi_{ij}=\hat\phi_{ij}/(1-\lam_{ij})$. Inserting into the constraints yields
\begin{equation}
  \frac{\hat\phi_{ij}}{1-\lam_{ij}} = \frac{\hat\phi_{ji}}{1+\lam_{ij}}.
\end{equation}
Solving for $\lam_{ij}$, we thus obtain the optimal fluxes
\begin{equation}
  \phi_{ij} = \frac{1}{2}(\hat\phi_{ij}+\hat\phi_{ji}),
\end{equation}
which are manifestly symmetric (and, in hindsight, not very surprising).

We can go a step further and, in a non-equilibrium steady state, consider the local detailed balance condition Eq.~\eqref{eq:ldb}, which yields the optimal fluxes
\begin{equation}
  \phi_{ij} = (\hat\phi_{ij}+\hat\phi_{ji})\left[1+\frac{p_j}{p_i}e^{(G^0_j-G^0_i)-f_\al d^\al_{ij}}\right]^{-1}.
\end{equation}
We now need to know also the equilibrium free energies $G^0_i$ of mesostates (which can be obtained from a separate equilibrium simulation using the same discretization) and the couplings $d^\al_{ij}$ together with the driving affinities.

%% ---- discussion ----

\section{Challenges}
\label{sec:disc}

\subsection{Variational principles: Caliber, transferability, and numerical sampling}
\label{sec:cal}

The task of obtaining transition rates for a set of mesostates is also addressed by the method of \emph{maximum caliber}~\cite{jayn80,presse13,dixit18}, which posits that the actual process maximizes the ``caliber'' or, equivalently, minimizes the distance between path weights as measured by the Kullback-Leibler divergence (cf. appendix~\ref{sec:kl}). Here the distance is with respect to a reference process $\tilde w$, whereby the minimization 
\begin{equation}
  \inf_{p,\phi} \bset{\mathcal I_{\tilde w}(p,\phi)}{\text{constraints}}
  \label{eq:cal}
\end{equation}
is constrained by the normalization of probabilities and available data about the process such as the average macroscopic currents $\mean{\mathcal J^\al}$. Introducing again Lagrange multipliers $\psi$, $\lam_{ji}=-\lam_{ij}$, and $\zeta_\al$, we thus have to minimize the function
\begin{multline}
  \mathcal I_{\tilde w}(p,\phi) - \frac{\psi}{2}\left(\sum_ip_i-1\right) - \frac{1}{2}\sum_{ij}\lam_{ij}(\phi_{ij}-\phi_{ji}-j_{ij}) \\ - \zeta_\al\left(\sum_{ij}d^\al_{ij}\phi_{ij}-\mean{\mathcal J^\al}\right).
\end{multline}
First, taking the derivative with respect to $p_i$ and setting it zero yields the condition
\begin{equation}
  \label{eq:cal:r}
  r_i = \tilde r_i - \frac{\psi}{2}
\end{equation}
between the exit rates of target and reference process. Second, taking the derivative with respect to the fluxes $\phi_{ij}$ then yields the rates
\begin{equation}
  \label{eq:cal:w}
  w_{ij} = \tilde w_{ij}e^{\lam_{ij}+\zeta_\al d^\al_{ij}}.
\end{equation}
This expression recovers a result obtained by Baule and Evans~\cite{baul08} following a different route (as noted previously~\cite{monthus11,chetrite15}), wherein rates obey the product constraint $w_{ij}w_{ji}=\tilde w_{ij}\tilde w_{ji}$, and Eq.~\eqref{eq:cal:r} is the exit rate constraint.

The most obvious choice for the reference process is thermal equilibrium with the affinities set to values so that there are no currents between reservoirs. We then have to tune the antisymmetric Lagrange multipliers $\lam_{ij}$ to fulfill the conditions Eq.~\eqref{eq:cal:r}. However, in general this tuning is not compatible with the local detailed balance condition since Eq.~\eqref{eq:cal:w} then implies
\begin{equation}
  \label{eq:cal:ldb}
  \ln\frac{w_{ij}}{w_{ji}} = -(G^0_j-G^0_i) + 2\lam_{ij} + 2\zeta_\al d^\al_{ij},
\end{equation}
which disagrees with Eq.~\eqref{eq:ldb} except for the special cases $\lam_{ij}=0$ and $\lam_{ij}\propto d^\al_{ij}$. While the resulting Markov model $w$ is driven and exhibits the right macroscopic currents (and entropy production $\mathcal S_\tau\asymp2\zeta_\al\Delta^\al_\tau$), these currents do not obey the constraints derived in Sec.~\ref{sec:ldb} due to physical exchanges with reservoirs. This failure points to a more fundamental issue, namely the existence of general variational principles out of equilibrium that fill in missing microscopic information comparable to the maximum entropy principle (see, \emph{e.g.}, Ref.~\citenum{polettini13} and Landauer's ``blow torch'' theorem~\cite{land75,land93}). Of course, one could constrain the minimization further and further until the desired result is obtained, but this somewhat defies the purpose of a variational principle. Moreover, on the practical side it limits this approach to rather simple systems with the number of mesostates comparable to the number of constraints~\cite{otten10}.

In fact, there is an exact variational principle if we rephrase the problem in terms of a \emph{biased} ensemble of trajectories~\cite{chet13,chet14,chetrite15}. The large deviation function $\psi(\zeta)$ is defined through $e^{-\tau\psi(\zeta)}\asymp\mean{e^{\zeta_\al\Delta^\al_\tau}}$, where the average is with respect to the original dynamics $w$. Now consider biasing the trajectories by changing their weight so that
\begin{equation}
  \mean{\Delta^\al_\tau}^\ast = \frac{\mean{\Delta^\al_\tau e^{\zeta_\al\Delta^\al_\tau}}}{\mean{e^{\zeta_\al\Delta^\al_\tau}}} = -\tau\pd{\psi}{\zeta_\al}.
\end{equation}
Tuning the parameters $\zeta$ thus promotes trajectories that have as typical exchanges $\mean{\Delta_\tau}^\ast$. This approach has been employed to study dynamic phase transitions~\cite{jack20} in the Ising model~\cite{jack10a,guioth20}, glasses~\cite{hedg09,garr09,spec12,campo20}, quantum systems~\cite{garr10}, and to extract nonlinear transport coefficients~\cite{gao19}. Since rate functions encode rare fluctuations they are difficult to compute directly in numerical simulations. To access extremely rare events that do not occur spontaneously on timescales accessible to the simulation, methods like transition path sampling~\cite{ray18}, population dynamics~\cite{giardina06,leco07,nemoto16}, and adaptive sampling~\cite{ferre18} have been developed.

A different approach to calcuating large deviations is to ask for a Markov model $w^\ast$ (in this context also called a control process) that exhibits as typical trajectories those with exchanges $\mean{\Delta_\tau}^\ast$ for given $\zeta$. This control process can now be found through minimizing (derived in appendix~\ref{sec:con})
\begin{equation}
  \inf_{p^\ast,\phi^\ast}\left\{\mathcal I_w(p^\ast,\phi^\ast)-\zeta_\al\sum_{ij}d^\al_{ij}\phi^\ast_{ij}\right\}.
  \label{eq:con}
\end{equation}
While similiar to Eq.~\eqref{eq:cal} on first glance~\cite{chetrite15}, this variational principle is conceptual different as it determines an auxillary control process that is not restricted, \emph{e.g.}, by local detailed balance. Note that this control process looks very different from another process $w'$ in which we only tune the affinities $f$ but keep the interparticle forces to achieve the same currents~\cite{spec16b}. In particular, the control process $w^\ast$ introduces long-range and many-body interactions and thus modifies the underlying atomistic model (cf. appendix~\ref{sec:bias}). This insight points to a major challenge that will have to be addressed, namely that of \emph{transferability}: Can we predict how the transition rates change as we change affinities without repeating the molecular dynamics simulations? As we have seen, local detailed balance [Eq.~\eqref{eq:ldb}] constrains the rates and determines the antisymmetric edge affinities $a_{ij}$. Writing the fluxes as $\phi_{ij}=\phi^0_{ij}e^{a_{ij}/2}$, it does not constraint the symmetric part $\phi^0_{ij}=\phi^0_{ji}$, which leaves considerable freedom~\cite{zia07}. First insights into how driving beyond linear response affects the time-symmetric action have been reviewed in Ref.~\citenum{maes20}.

The long-range and many-body nature of interactions in the auxillary control process makes the optimization Eq.~\eqref{eq:con} challenging. Recently, for continuous configuration spaces this has become an active field developing advanced numerical methods including parametric representations~\cite{das19}, reinforcement learning~\cite{rose21}, as well as other machine learning techniques~\cite{oakes20,yan21}.

\subsection{Coarse-graining and metastable basins}

\begin{figure}[b!]
  \centering
  \includegraphics{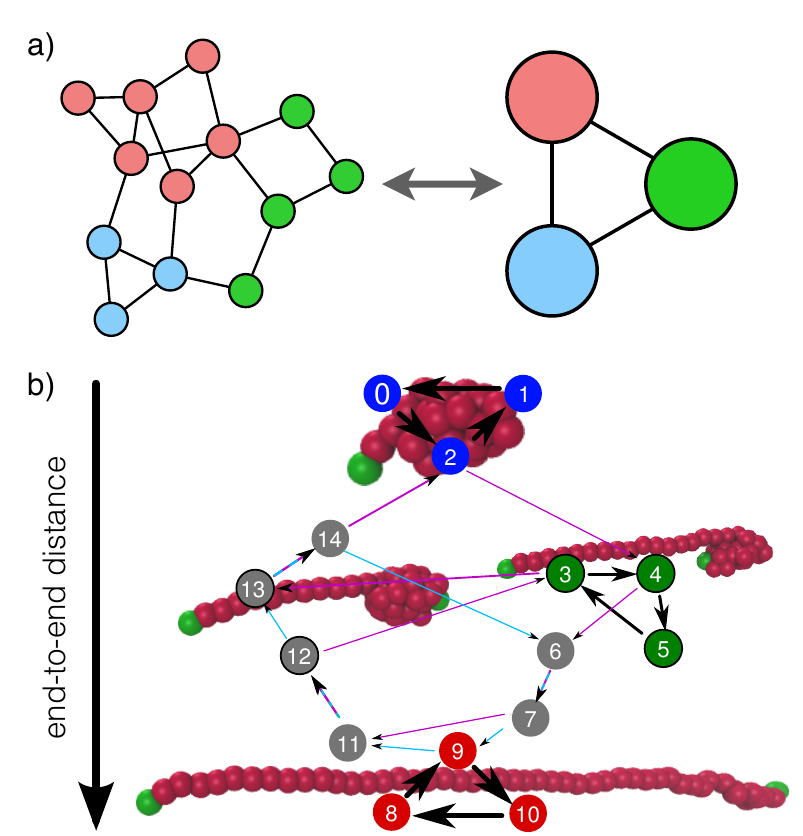}
  \caption{(a)~Coarse-graining of mesostates (left) into a few basins (right) corresponding to long-lived structural states. (b)~Specific example of a grafted polymer in shear flow~\cite{knoc17}. Shown is the minimal graph of mesostates together with representative snapshots of the polymer. Clustering of cycles reveals five communities, three of which are local corresponding to dynamics with small fluctuations of the end-to-end distance [globule (blue), stretched (red), and one intermediate (green)], and two global communities cycling through the collapse and stretching of the polymer.}
  \label{fig:meta}
\end{figure}

While here we have discussed continuous-time Markov processes with transition rates $w$, conventional Markov state modeling is rather build around Markov chains characterized by the transition matrix $T$, the entries $T_{ij}$ of which are the probabilities to observe a transition $i\to j$ in a fixed time window $\delta t$. Detailed balance guarantees that all eigenvalues of this matrix are real and positive. The largest eigenvalue is equal to unity and its eigenvector is composed of the steady-state probabilities $p$. Ordering eigenvalues from large (slow) to small (fast), the following eigenvectors describe the large-scale dynamics. A gap in the spectrum of these eigenvalues indiates a further timescale separation and allows to identify metastable basins (disjoint sets of mesostates), cf. Fig.~\ref{fig:meta}(a). These basins correspond to long-lived, in the case of proteins and peptides often misfolded, structural states. It is then straightforward to construct another Markov process between these basins.

In a non-equilibrium steady state, the largest eigenvalue of the transition matrix is still unity and non-degenerate, but other eigenvalues may become complex (these appear in complex conjugate pairs). More importantly, there are now non-vanishing probability currents coupling to macroscopic currents. Simply combining mesostates into superstates will affect these currents and reduce the entropy production~\cite{pugl10}. The aim is thus to derive a Markov process with a reduced number of states that yields the same (average) macroscopic currents, and thus entropy production.

In fact, this goal can be achieved rigourously~\cite{knoc15}. To this end, we focus on the cycles in the graph of mesostates. As shown in appendix~\ref{sec:decomp}, the probability currents
\begin{equation}
  \label{eq:j:de}
  j_e = \sum_c \vhi_c c_e
\end{equation}
can be decomposed into contributions from cycles, where $\vhi_c\geqslant 0$ is the weight of cycle $\vec c=(c_e)$. While the number of cycles can be very large, the number of contributing cycles with $\vhi_c>0$ is bound by the Betti number $|E|-|V|+1$. Note that the weights $\vhi_c$ do not have a physical meaning but allow to reconstruct currents from a set of contributing cycles.

Conjugate to the edge currents $j_e$ are the edge affinities $a^e=a_{ij}$ for each $e=i\to j$. Plugging the decomposition Eq.~\eqref{eq:j:de} into the entropy production rate Eq.~\eqref{eq:sig}, we obtain $\dot\sig=\sum_e a^ej_e=\sum_c\vhi_ca(\vec c)$ with cycle affinities
\begin{equation}
  a(\vec c) \equiv \sum_e a^ec_e = f_\al\sum_{(i\to j)\in\vec c} d^\al_{ij} = f_\al d^\al(\vec c),
\end{equation}
which are given by the net transport between reservoirs (we sum over all directed edges that form the cycle). The next step is to decimate the number of contributing cycles. This is made possible by realizing that the distribution of cycles is far from uniform as many cycles visit neighboring mesostates and thus cluster into communities, which can be identified by well established techniques. Let us assume that we have obtained such a partition into sets $\{\mathcal C_k\}$ of cycles. We pick one representative cycle $k\in\mathcal C_k$ for each community. Which cycle is not important for what follows, and representatives can be choosen to be optimal in some sense (\emph{e.g.}, optimize the mean-first passage times between communities~\cite{knoc17}). The linearity of the entropy production rate is exploited to split
\begin{equation}
  \dot\sig = \sum_k \dot\sig_k, \qquad \dot\sig_k \equiv \sum_{c\in\mathcal C_k} \vhi_c a(\vec c)
\end{equation}
into a contribution from each community. Eliminating all cycles except for the representative, it has to carry all the entropy production, which implies the new weight $\vhi'_k=\dot\sig_k/a_k$. As a consequence, the new currents are $j'_e=\sum_k\vhi'_kk_e$.  However, we still have to determine the symmetric part of the fluxes as well as the probabilities. We know that the local detailed balance condition Eq.~\eqref{eq:ldb} fixes the ratio of the transition rates, which can be ensured by fixing the ratio $p'_i/p'_j=p_i/p_j$ and preserving edge affinities
\begin{equation}
  e^{a_{ij}} = \frac{\phi_{ij}}{\phi_{ji}} = \frac{\phi'_{ij}}{\phi'_{ji}} = \frac{\phi'_{ji}+j'_{ij}}{\phi'_{ji}}.
\end{equation}
This condition allows us to determine the new fluxes
\begin{equation}
  \phi'_{ij} = \frac{e^{a_{ij}}}{e^{a_{ij}}-1}\sum_k\vhi'_kk_{ij}
\end{equation}
with $k_{ij}=\pm 1$ if the oriented edge $i\to j$ participates in cycle $\vec k$. We have thus derived a Markov process with transition rates $w'_{ij}=\phi'_{ij}/p_i$ between the mesostates that are included in one of the representative cycles. This Markov process rigorously preserves the entropy production and the macroscopic currents between reservoirs.

As an illustration, Fig.~\ref{fig:meta}(b) shows the result of this procedure~\cite{knoc17} for a grafted model polymer in shear flow~\cite{alex06}. For intermediate strain rates, the polymer cycles between globule and extended conformations. A fine-grained Markov model has been obtained through clustering structurally similar configurations harvested from simulations, from which the rates $w$ are obtained through counting. Determining all contributing cycles in this fine-grained model reveals five cycle communities largely determined by the visited end-to-end distances, cf. Fig.~\ref{fig:meta}(b). The minimal coarse-grained model then consists of only 15 mesostates after eliminating further ``bridge states'' that are unique to their cycles (see also Ref.~\cite{seiferth20}). Periodically driven molecules (\emph{e.g.}, due to the coupling of a residual dipole moment with an external electric field) have also been addressed~\cite{wang15,knoc19}.

%% ---- conclusions ----

\section{Conclusions and outlook}

Markov state modeling and dimensional reduction techniques~\cite{pand10,prin11,sittel18} have been successfully employed to extract the dynamics of discrete molecular conformations from atomistic models of biomolecules, primarily peptides. Analyzing these Markov models yields a comprehensive picture of possible (\emph{e.g.}, folding) pathways and kinetics. Extending this framework to molecular machines requires to take into account the entropy that is dissipated into the aqueous environment. Splitting the total system into the molecular motor itself and idealized reservoirs [cf. Fig.~\ref{fig:system}(a)] has the advantage that the entropy production is determined completely by the exchange with the reservoirs [Eq.~\eqref{eq:diss}]. These, typically few, currents take the role of extensive variables in a framework formally equivalent to thermodynamics and based on large deviation theory. Preserving these currents allows to systematically construct a coarse-grained representation of the internal dynamics in terms of molecular conformations.

On the simplest level discussed here, equilibrium detailed balance is replaced by the local detailed balance condition Eq.~\eqref{eq:ldb} modeling exchanges with ideal reservoirs. While clearly being a simplification in most situations, the advantage is that we do not have to take these reservoirs into account explicitly. Moreover, as discussed in some detail here, it allows a consistent view on the stochastic energetics and thermodynamics. What we need in addition to equilibrium Markov state models is to identify which conformational transitions involve exchanges with the reservoirs, which requires some physical insight into these conformations. However, one might expect that the matrices $d^\al_{ij}$ are sparse with only very few transitions being involved. For example, ATP might only bind to a single conformation. Another computational advantage is that we do not need to resolve the quantum-mechanical nature of this binding transition, while thermodynamic consistency is maintained by construction.

There have been tremendous advances, not only in our grasp of biological, but also in the design and synthesis of artificial molecular machines~\cite{erba15,kistemaker21}. Modeling will play an increasing role but requires robust methods based on rigorous principles out of equilibrium. Here we have sketched how to deploy stochastic thermodynamics for the systematic construction of coarse-grained models, which will complement more detailed atomistic inverstigations.

%% ---- acknowledgments ----

\begin{acknowledgments}
  I acknowledge financial support by the Deutsche Forschungsgemeinschaft through the TRR 146 ``Multiscale Simulation Methods for Soft Matter Systems'' (project A7). I thank Fabian Knoch for many useful discussions as well as Udo Seifert for his inspiration and guidance.
\end{acknowledgments}

%% ---- appendix ----

\appendix

\section{Derivation of Rayleighian}
\label{sec:ray}

To lowest order in the linear response regime around equilibrium, the generating function reads $\psi(f)=-\frac{1}{2}f_\al\chi^{\al\beta}f_\beta$ in agreement with Eq.~\eqref{eq:chi}. This is a convex function, and we can thus determine its inverse Legendre transformation (as a functional of $\dot X_t$),
\begin{align*}
  \tau\Phi[\dot X_t] &= \sup_f[f_\al\Delta^\al+\tau\psi(f)] \\ 
  &= \sup_f \IInt{t}{0}{\tau}[f_\al\dot X_t^\al-\tfrac{1}{2}f_\al\chi^{\al\beta}f_\beta] \\
  &= \IInt{t}{0}{\tau} \sup_f[f_\al\dot X_t^\al-\tfrac{1}{2}f_\al\chi^{\al\beta}f_\beta] \\
  &= \IInt{t}{0}{\tau} \tfrac{1}{2}\dot X_t^\al\chi_{\al\beta}\dot X_t^\beta.
\end{align*}
Here, we multiply by the trajectory length $\tau>0$, pull out the integral using Eq.~\eqref{eq:Delta}, and exploit that we can interchange the supremum with the integration~\cite{hafsa03}. Multiplying the right hand side of Eq.~\eqref{eq:onsa:psi} by $\tau$ and inserting the result for $\tau\Phi$, we again interchange
\begin{multline}
  \inf_{\dot X_t} \IInt{t}{0}{\tau}[\tfrac{1}{2}\dot X_t^\al\chi_{\al\beta}\dot X_t^\beta-f_\al\dot X_t^\al] \\ = \IInt{t}{0}{\tau}\inf_{\dot X}[\tfrac{1}{2}\dot X^\al\chi_{\al\beta}\dot X^\beta-f_\al\dot X^\al]
\end{multline}
yielding $\inf_{\dot X}\mathcal R$ with the function $\mathcal R$ given in Eq.~\eqref{eq:ray}.

\section{Cycle space}
\label{sec:cyc}

To see that the null space of the incidence matrix $\nabla$ is the space of directed simple \emph{cycles} pick a starting vertex $i_0$ and an edge $e$ connecting it to $i_1$. For the vector $\vec c$ set $c_e=+1$ if the edge enters $i_1$ and $c_e=-1$ if it leaves $i_1$ (all other entries are zero). Applying the incidence matrix then leads to a vector with entry $+1$ for $i_1$ and $-1$ for $i_0$. Pick a second edge connecting $i_1$ with $i_2$ and set $c_e=\pm1$ depending on the edge orientation. Again applying the incidence matrix, the entry for $i_1$ has become zero and the entry for $i_2$ is $+1$. This construction can be continued until the last edge enters the starting vertex $i_0$. Now the vector $\nabla\cdot\vec c=\nabla_i^ec_e=0$ has all entries zeros and we stop. Hence, a (simple) cycle $i_0\to i_1\to\cdots\to i_0$ in which all visited vertices are unique can be represented as an edge vector with entries $c_e=\pm1$ for edges that are part of the cycle and $c_e=0$ for edges that are not part of the cycle. The sign is positive if the cycle edge has the same orientation as the graph edge.

\section{Cycle decomposition}
\label{sec:decomp}

\begin{figure}[b!]
  \centering
  \includegraphics{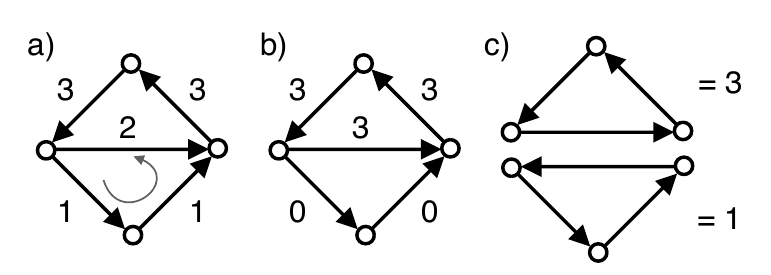}
  \caption{Illustration of the cycle decomposition. (a)~Simple directed graph $\vec G$ with four vertices. The numbers indicate the current along the corresponding edge. Note that the current law Eq.~(\ref{eq:kirch}) is obeyed. We first consider the cycle through the lower three states (arrow) with smallest current $j_e=1$. Subtracting this current from (or, if antiparallel, adding to) all edges of this cycle leads to (b). (c)~We thus find two contributing cycles with weights $\vhi_1=3$ (top) and $\vhi_2=1$ (bottom) that share one edge.}
  \label{fig:decomp}
\end{figure}

The following algorithm can be used to determine the cycle weights $\vhi_c$ iteratively~\cite{kalpazidou07,alta12a}. We enumerate all cycles starting with the first cycle, $c=1$, and we initialize $j_e^{(1)}=j_e$ with the currents. We go along all edges of the cycle with $c_e=+1$ (parallel to graph edge) and pick the smallest current, which we assign as cycle weight,
\begin{equation}
  \vhi_c = \min_{e\in E}\{j_e^{(c)}(\vec c)|c_e=+1\} \geqslant 0.
\end{equation}
Remember that $j_e\geqslant 0$ by construction and thus also the cycle weight is non-negative. We subtract this cycle weight from all cycle edges,
\begin{equation}
  j_e^{(c+1)} = j_e^{(c)} - \vhi_c c_e.
\end{equation}
Either $\vhi_c=0$ (because one or more edges had zero current) and nothing changes, or the new current vector $j_e^{(c+1)}$ has one more entry equal to zero with all other entries still non-negative. Note that $\sum_e\nabla_i^ej_e^{(c+1)}=0$. We then go to the next cycle $c\leftarrow c+1$ and repeat this procedure. The iteration is stopped when all entries $j_e^{(c+1)}$ are zero and thus all contributing cycles have been identified. Fig.~\ref{fig:decomp} illustrates this procedure for a simple system with four mesostates.

\section{Kullback-Leibler divergence}
\label{sec:kl}

The Kullback-Leibler divergence (or relative entropy) is often cited to determine a ``distance'' between probability measures. In case of trajectories, it can be written
\begin{equation}
  \label{eq:kl}
  \mathcal D_\text{KL}(\mathcal P^\ast_\tau\parallel\mathcal P_\tau) \equiv \sum_{\trj} \mathcal P^\ast_\tau[\trj]\ln\frac{\mathcal P^\ast_\tau[\trj]}{\mathcal P_\tau[\trj]} = \tau\mathcal I_w(p^\ast,\phi^\ast),
\end{equation}
relating it to the rate function Eq.~\eqref{eq:rf}. The final result follows from plugging in the trajectory weight Eq.~\eqref{eq:traj} for $\mathcal P_\tau$, and the same expression for $\mathcal P^\ast_\tau$ but replacing $w\to w^\ast$. For the logarithmic ratio we obtain (ignoring boundary terms)
\begin{equation}
  \sum_{\nu=1}^K \ln\frac{w^\ast_{i_{\nu-1}i_\nu}}{w_{i_{\nu-1}i_\nu}} - \sum_{\nu=0}^K \sum_j (w^\ast_{i_\nu j}-w_{i_\nu j})(t_{\nu+1}-t_\nu).
\end{equation}
Taking the average and dividing by $\tau$ then yields
\begin{equation}
  \sum_{ij}\phi^\ast_{ij}\ln\frac{w^\ast_{ij}}{w_{ij}} - \sum_{ij}p^\ast_i(w^\ast_{ij}-w_{ij}) = \mathcal I_w(p^\ast,\phi^\ast),
\end{equation}
namely the rate function Eq.~\eqref{eq:rf} determining the probability to jointly observe populations $p^\ast$ and fluxes $\phi^\ast$.

\section{Optimal control process}
\label{sec:con}

To derive Eq.~\eqref{eq:con}, we start from the large deviation function
\begin{equation}
  -\tau\psi \asymp \ln\mean{e^{\zeta_\al\Delta^\al_\tau}} = \ln\sum_{\trj} e^{\zeta_\al\Delta^\al_\tau} \frac{\mathcal P_\tau[\trj]}{\mathcal P^\ast_\tau[\trj]}\mathcal P^\ast_\tau[\trj]
\end{equation}
and expand the path weight by the weight $\mathcal P^\ast_\tau[\trj]$ of the (unknown) control process. In the limit of large $\tau$, the right hand side is dominated by the saddle-point leading to
\begin{equation}
  \ln\left\langle\exp\left\{\zeta_\al\Delta^\al_\tau+\ln\frac{\mathcal P_\tau}{\mathcal P^\ast_\tau}\right\}\right\rangle^\ast \asymp \sup_{w^\ast}\left\langle \zeta_\al\Delta^\al_\tau+\ln\frac{\mathcal P_\tau}{\mathcal P^\ast_\tau} \right\rangle^\ast
\end{equation}
with respect to all admissible Markov processes $w^\ast$. Performing the average and using Eq.~\eqref{eq:kl}, we thus find
\begin{equation}
  \tau\sup_{w^\ast}\left\{\zeta_\al\mean{\mathcal J^\al}^\ast-\mathcal I_w(p^\ast,\phi^\ast)\right\},
\end{equation}
which yields Eq.~\eqref{eq:con} after pulling out the minus sign.

\section{Biasing potential}
\label{sec:bias}

To see how the biasing of trajectories changes the underlying free energy landscape, we follow Ref.~\citenum{spec11}. We plug the rates Eq.~\eqref{eq:cal:w} together with Eq.~\eqref{eq:cal:r} into the path weight Eq.~\eqref{eq:traj},
\begin{equation}
  \mathcal P_\tau[\trj] = \tilde{\mathcal P}_\tau[\trj] \exp\left\{\zeta_\al\Delta^\al_\tau+\tau\frac{\psi}{2}+\sum_{\nu=1}^K\lam_{i_{\nu-1}i_\nu}\right\},
\end{equation}
which thus can be written as the original path weight times an exponential factor. Summing over all trajectories and imposing normalization of the biased ensemble yields
\begin{equation}
  e^{-\tau\psi(\zeta)/2} = \left\langle\exp\left\{\zeta_\al\Delta^\al_\tau+\sum_{\nu=1}^K\lam_{i_{\nu-1}i_\nu}\right\}\right\rangle_\text{eq},
\end{equation}
with $\psi(\zeta)$ the large deviation function.

For the antisymmetric Lagrange multipliers $\lam_{ij}$ we now write $\lam_{ij}=(v_i-v_j)/2$. Plugging this expression into the ratio Eq.~\eqref{eq:cal:ldb} and comparing with the local detailed balance condition Eq.~\eqref{eq:ldb}, we see that $\zeta_\al=f_\al/2$ can be identified with the affinities but the free energies are shifted as $G^0_i\to G^0_i+v_i$. Hence, the $v_i(\zeta)$ constitute a biasing potential that modifies the underlying atomistic model at variance with the local detailed balance condition Eq.~\eqref{eq:ldb}. The sum
\begin{equation}
  \sum_{\nu=1}^K\lam_{i_{\nu-1}i_\nu} = v_{i_0} - v_{i_K}
\end{equation}
reduces to a temporal boundary term that does not change $\psi(\zeta)$. Using that the sum $\sum_{(i\to j)\in c}\lam_{ij}=0$ along any simple cycle $c$ vanishes together with the exit rate constraints [Eq.~\eqref{eq:cal:r}] allows (in principle) to determine $v$ and $\psi$.

For affinities corresponding to thermal equilibrium, we have $f_\al\Delta^\al=0$ and $\psi=0$ as well as $v=0$. Close to equilibrium, we expand
\begin{equation}
  e^{-\tau\psi(f)/2} = \left\langle e^{\tfrac{1}{2}f_\al\Delta^\al_\tau}\right\rangle_\text{eq} + \mathcal O(v^2)
\end{equation}
since $\mean{(v_i-v_j)}=0$. Hence, for a small perturbation of equilibrium no shifts $v$ occur, which justifies the ansatz Eq.~\eqref{eq:onsa:P} for the probability density of the changes $\Delta^\al_\tau$ in the linear response regime.

%% ---- bibliography ----


\end{document}